# A statistical thin-tail test of predicting regulatory regions in the *Drosophila* genome

Jian-Jun SHU[1] and Yajing LI

School of Mechanical & Aerospace Engineering, Nanyang Technological University, 50 Nanyang Avenue, Singapore 639798

## Abstract

**Background**

The identification of transcription factor binding sites (TFBSs) and *cis*-regulatory modules (CRMs) is a crucial step in studying gene expression, but the computational method attempting to distinguish CRMs from NCNRs still remains a challenging problem due to the limited knowledge of specific interactions involved.

**Methods**

The statistical properties of *cis*-regulatory modules (CRMs) are explored by estimating the similar-word set distribution with overrepresentation (*Z-score*). It is observed that CRMs tend to have a thin-tail *Z-score* distribution. A new statistical thin-tail test with two thinness coefficients is proposed to distinguish CRMs from non-coding non-regulatory regions (NCNRs).

---

[1] Correspondence should be addressed to Jian-Jun SHU, mjjshu@ntu.edu.sg

**Results**

As compared with the existing fluffy-tail test, the first thinness coefficient is designed to reduce computational time, making the novel thin-tail test very suitable for long sequences and large database analysis in the post-genome time and the second one to improve the separation accuracy between CRMs and NCNRs. These two thinness coefficients may serve as valuable filtering indexes to predict CRMs experimentally.

**Conclusions**

The novel thin-tail test provides an efficient and effective means for distinguishing CRMs from NCNRs based on the specific statistical properties of CRMs and can guide future experiments aimed at finding new CRMs in the post-genome time.

# 1. Background

The identification of transcription factor binding sites (TFBSs) and *cis*-regulatory modules (CRMs) is a crucial step in studying gene expression. The computational methods of predicting CRMs from non-coding non-regulatory regions (NCNRs) can be classified into three types: 1) TFBS-based methods, 2) homology-based methods and 3) content-based methods. TFBS-based methods, such as ClusterBuster [1] and MCAST [2], use information about known TFBSs to identify potential CRMs. The methods of this type are limited to the recognition of similarly regulated CRMs, and generally unable to be applied to genes for which TFBSs have not yet been studied experimentally. Homology-based methods use information contained in the pattern of conservation among related sequences. The related sequences can come from single species [3], two species [4] and multiple species [5]. The methods of this type using the pattern of conservation alone are limited in their performance because TFBS conservation necessary to maintain regulatory function in binding sequences may not be significantly higher than in non-binding sequences [6,7]. In addition, it still remains an open question that how many genomes are sufficient to the reliable extraction of regulatory regions. Content-based methods assume that different genome regions (CRMs and NCNRs) have the different rates of evolutionary micro changes; therefore, they exhibit different statistical properties in nucleotide composition. TFBSs often occur together in clusters as CRMs. The binding site cluster causes a biased word distribution within CRMs, and this bias leaves a distinct "signature" in nucleotide composition. Content-based methods detect this signature by statistical [8,9] or machine-learning [10,11] techniques, in order to distinguish CRMs from non-CRMs. The methods of this type may be used to predict the CRMs which have not yet been observed experimentally, but the poor performance on non-coding sequences limits their applications [12]. A large number of CRM search tools have been reported in the literature, but the computational method attempting to distinguish CRMs from NCNRs still remains a challenging problem due to the limited knowledge of specific interactions involved [13].

The fluffy-tail test [9] is one of content-based methods. It is a bootstrapping procedure to recognize statistically significant abundant similar-words in CRMs. There are two problems with the fluffy-tail test: 1) Due to its bootstrapping procedure, the computational time of calculating the fluffiness coefficient is determined by the number of realization. In order to get reliable results statistically, the number of realization is usually set as very large in the fluffy-tail test, so the computational time is expensive, especially for long sequences. This limits the use of the fluffy-tail test under the situation when more and more DNA sequences need to be analyzed in the post-genome time. 2) The separation performance between CRMs and NCNRs is far from satisfactory [12]. The reason of poor performance is that both CRMs and NCNRs contain repetitive elements such as poly(N) tracts (… TTT…) or long simple repeats (…CACACA…). These strings are less interesting than the over-represented strings with more balanced AT/GC ratio. It is an interest to address these two issues of the fluffy-tail test and to develop a more efficient and effective CRM prediction method.

In this paper, the statistical properties of CRMs are explored by evaluating the overrepresentation value of similar-word sets (motifs). *Z-score* is used as the measure of overrepresentation of similar-word sets. Then, *Z-score* distribution is estimated to distinguish CRMs from NCNRs.

## 2. Methods

### 2.1 Training datasets

To estimate the statistical properties of distinguishing CRMs from NCNRs, two (positive and negative) training datasets are employed in this paper. The positive training dataset is a collection of 60 experimentally-verified functional *Drosophila melanogaster* regulatory regions [14]. The positive training dataset consists of CRMs located far from gene coding sequences and transcription start sites. It contains many binding sites and site clusters, including *abdominal-b*, *bicoid*, *caudal*, *deformed*, *distal-less*, *engrailed*, *even-skipped*, *fushi tarazu*, *giant*, *hairy*, *huckebein*, *hunchback*, *knirps*, *krüppel*, *odd-paired*, *pleiohomeotic*, *runt*, *tailless*, *tramtrack*, *twist*, *wingless* and *zeste*. The total size of the positive training dataset comprises about 99 kilobase (kb) sequences. The negative training dataset is 60 randomly-picked *Drosophila melanogaster* NCNRs: The NCNRs of length 1 kb upstream and downstream of genes are excluded by using the Ensembl genome browser. The negative training dataset contains 90 kb sequences in total.

### 2.2 Formulation of the thin-tail test

The thin-tail test is based on the assumption that each word (binding site) recognized by a given transcription factor belongs to its own family of similar-word sets (binding site motifs) found in the same enhancer sequence and the redundancy of the binding sites within CRMs leaves distinct "signatures" in similar-word set distribution. For a given $m$-letter segment $W_m$ as a seed-word, all $m$-letter words that differ from $W_m$ by no more than $j$ substitution comprise a corresponding similar-word set $N_j(W_m)$. Because the core of TFBSs is relatively short [15], a 5-letter seed-word is considered, allowing for 1 mismatch, *i.e.*, $m = 5$ and $j = 1$. In order to distinguish CRMs from NCNRs, the thin-tail test is adopted to study the *Z-score* distribution shape and to predict the probable function of the original input sequence. The test features special statistics accounting for word overlaps in the same DNA strand. A flow chart of the thin-tail test is shown in Figure 1.

Step 1: Search for all different seed-words ($W_m$)

The input sequence is scanned to find all the different $m$-letter words, allowing overlaps. As an example, consider a stretch of DNA: ACGACGCCGACT. For $m = 5$, all 5-letter segments $W_5$ are selected as seed-words, *i.e.*, ACGAC, CGACG, …, CGACT.

Step 2: Number of similar-words with the same seed-word ($n$)

For each seed-word $W_m$, all $m$-letter words with no more than $j$ substitution comprise a corresponding similar-word set $N_j(W_m)$. In this example, the first seed-word $W_5$, ACGAC, has 3 similar-words with no more than 1 mismatch: ACGAC, ACGCC, CCGAC. $n$ is the cardinality, $n = |N_j(W_m)| = |N_1(\text{ACGAC})| = 3$.

Step 3: *Z-score* with the same seed-word ($Z$)

A similar-word set that occurs significantly more often than chance expectation is said to be overrepresentation. A reasonable overrepresentation measure would reflect whether the actual occurrence number of similar-word set is significantly greater than the number counted in a random sequence with the same composition of input sequence. For any seed-word $W_m$, a statistical overrepresentation measure *Z-score* can be defined by

$$Z = \frac{n - E(W_m)}{\sqrt{V(W_m)}} \quad (1)$$

where $E(W_m)$ and $V(W_m)$ are, respectively, the occurrence expectation and variance of similar-word set $N_j(W_m)$, these being calculated for a random sequence with the same composition of input sequence [16]. In a random Bernoulli type sequence, both occurrence expectation and variance can be derived analytically by using a generating function technique [17]. The *Z-score* with more overlaps is smaller than one with less overlaps. For example, the *Z-score* corresponding to simple repeat strings, TTTTT or AAAAA, is smaller than one corresponding to the seed-word with more balanced composition. $Z$ (*Z-score*) forms X axis in Figures 2-5.

Step 4: Number of seed-words with the same *Z-score* ( $f$ )

$f(Z)$ is the number of the seed-words with *Z-score* and forms Y axis in Figures 2-7.

Step 5: Kurtosis ( $k$ )

The kurtosis $k$ of *Z-score* distribution $f(Z)$ is evaluated as

$$k = \frac{\sum_{i=1}^{M}[f(Z)-\mu]^4}{(M-1)\sigma^4} - 3 \qquad (2)$$

where $i$ is the $i$ th seed-word, $M$ is the total number of seed-words, $\mu$ and $\sigma$ are the mean and standard deviation of *Z-score* distribution $f(Z)$ respectively.

Step 6: Two thinness coefficients ( $E$ and $T_r$ )

The first thinness coefficient $E$ is defined as:

$$E = \frac{k_0 + 2\varepsilon}{4\varepsilon}. \qquad (3)$$

Here $k_0$ denotes the kurtosis $k$ of the original input sequence without random shuffle and $\varepsilon$ is the standard error calculated by:

$$\varepsilon = 2\sqrt{\frac{6}{M}}. \qquad (4)$$

$E$ is used to measure how strongly *Z-score* distribution deviates from the normal distribution. The 95% confidence interval is set between $-2\varepsilon$ and $2\varepsilon$.

A sequence is called "random" if it is obtained by randomly shuffling the original input sequence $r$ times, preserving its single nucleotide composition. To measure how strongly the *Z-score* distribution deviates from randomness, the second thinness coefficient $T_r$ is computed by comparing with all $r$-times randomly-shuffled sequence versions of the original input sequence:

$$T_r = \frac{k_0 - k_r}{\sigma_r} \quad . \tag{5}$$

Here $T_r$ can be regarded as measuring the degree of the difference between signal and noise, where the signal is regarded as the original input sequence, and the noise is regarded as randomized sequences.

In the fluffy-tail test [9], the fluffiness coefficient $F_r$ is defined as:

$$F_r = \frac{L_0 - L_r}{s_r} \tag{6}$$

where $L_r$ is the number of the seed-words with the maximal similar-words for the $r$-times randomly-shuffled sequences and $s_r$ is the standard deviation of the similar-word set distribution between the number $g(n)$ of seed-words and the number $n$ of similar-words. Here it is worth to mention to this end that both CRMs and NCNRs contain repetitive elements such as poly(N) tracts (… TTT…) or long simple repeats (…CACACA…), which are less interesting than the over-represented strings with more balanced AT/GC ratio. Since *Z-score* measures the overrepresentation of similar-word sets, the second thinness coefficient $T_r$ based on *Z-score* distribution should be a more reasonable index than the fluffiness coefficient $F_r$ based on similar-word set distribution in order to distinguish CRMs from NCNRs.

## 3. Results

### 3.1 Distribution for CRMs

Figure 2 shows the *Z-score* distribution for all *Drosophila melanogaster* CRMs in the positive training dataset. It can be seen that some similar-word sets have extreme positive/negative *Z-score* ($Z > 3$ or $Z < -3$). This means that some similar-word sets are overrepresented or underrepresented.

To obtain a random distribution, the original sequence is randomly shuffled $r = 50$ times. Figure 3 shows a typical example of *Z-score* distribution after random shuffle. As compared with the original input sequence in Figure 2, the randomized sequence in Figure 3 lacks the overrepresented/underrepresented similar-word set (*i.e.* similar-word set with extreme *Z-score*, $Z > 3$ or $Z < -3$).

### 3.2 Distribution for NCNRs

Figure 4 shows the *Z-score* distribution for all randomly-picked *Drosophila melanogaster* NCNRs in the negative training dataset. The presence of short right tail is noted in Figure 4. Figure 5 shows a typical example of *Z-score* distribution after random shuffle. The distribution for the original input sequence notably differs from that for the randomized version. The difference degree of the distribution between the original and randomly-shuffled sequences for NCNRs is greater than that for CRMs.

### 3.3 Thin-tail test

In order to distinguish CRMs from NCNRs, $E$ and $T_r$ are calculated for 120 sequences in these two training datasets. Figure 6 shows that CRMs tend to have a smaller $E$ than NCNRs. Table 1(a) lists functional classification based on $E$. Nearly 71.7% CRMs has $E < 0.6$, while only 41.7% NCNRs has $E < 0.6$. Figure 7 shows $T_{50}$ for CRMs and NCNRs. For each sequence, its *50*-times randomly-shuffled versions are generated to calculate $T_{50}$. It can be seen that CRMs tend to have a smaller $T_{50}$ than NCNRs. Table 1(b) lists functional classification based on $T_{50}$. Nearly 73.3% CRMs has $T_{50} < 0$, while only 40% NCNRs has $T_{50} < 0$.

## 4. Discussion

Some statistical properties of *Z-score* distribution in these two training datasets have been explored. Results show that CRMs have a thin-tail distribution, *i.e.*, tend to have low thinness coefficients ($E < 0.6$, $T_r < 0$), while NCNRs lack a thin-tail distribution, *i.e.*, tend to have high fatness coefficients. Thus, $E$ and $T_r$ can be used to distinguish CRMs from NCNRs effectively. CRMs are predominant if ($E < 0.6$, $T_r < 0$), while NCNRs are prevailing if ($E > 0.6$, $T_r > 0$). Thus, the regions with ($E < 0.6$, $T_r < 0$) are CRMs and those with ($E > 0.6$, $T_r > 0$) are NCNRs.

### 4.1 Comparison with fluffy-tail test

The thin-tail test is evaluated by comparison with the fluffy-tail test [9]. The performance of three parameters is assessed: 1) the first thinness coefficient $E$, 2) the second thinness coefficient $T_r$ and 3) the fluffiness coefficient $F_r$ based on the separation between CRMs and NCNRs.

These two training datasets are employed to evaluate the above three parameters. For comparison, the original input sequence is randomly shuffled *50* times to calculate $T_{50}$ and $F_{50}$. The thresholds of $E$, $T_{50}$ and $F_{50}$ are set as 0.6, 0 and 2 respectively. For the thin-tail test, the original input DNA sequence with $E < 0.6$ and $T_{50} < 0$ is considered as predicted CRMs. For the fluffy-tail test, the original input DNA sequence with $F_{50} > 2$ is considered as predicted CRMs. The classification result of 120 sequences in these two training datasets by $F_{50}$ is listed in Table 1(c). The fluffy-tail test $F_{50}$ only identified 29 out of 60 NCNRs in the negative training dataset; while the thin-tail test identified 35 and 36 NCNRs based on $E$ and $T_{50}$ respectively (see Table 1). For each parameter, sensitivity (SN) (number of true positive/number of positive), specificity (SP) (number of true negative/number of negative) and accuracy (number of true positive+number of true negative)/(number of positive+number of negative) are calculated to distinguish CRMs from NCNRs (Table 2).

The thin-tail test with $T_{50}$ has the best accuracy (66.7%), as compared with the other two parameters ($E$: 65%; $F_{50}$: 65%). Thus, the thin-tail test with $T_{50}$ can effectively distinguish CRMs from NCNRs. Moreover, the thin-tail test (SP = 60% for $T_{50}$ and SP = 58.3% for $E$) can more efficiently identify NCNRs than the fluffy-tail test (SP = 48.3% for $F_{50}$). The thin-tail test with $E$ has the same accuracy as the fluffy-tail test. However, the computational time (CPU time) of calculating $E$ for an original input DNA sequence length of 1000 is 50 times faster than that of calculating $T_{50}$ and 6 times faster than that of calculating $F_{50}$ for the same original input sequence due to no sequence shuffle. Thus, the thin-tail test with $E$ is very suitable for long sequences and large database.

### 4.2 Time complexity

The second thinness coefficient $T_r$ is gotten by bootstrapping procedure, the value is affected by the number of realization $r$. In order to get the more reliable estimation of $T_r$, a large $r$ is needed, so that high computational time is expected. For the reliable result within reasonable computational time, the original input sequence is randomly shuffled *50* times to calculate $T_r$.

In Table 2(c), the computational time (CPU time) of calculating $E$ for an original input DNA sequence length of 1000 is 50 times faster than that of calculating $T_{50}$ and 6 times faster than that of calculating $F_{50}$ for the same original input sequence due to no sequence shuffle. All computations are run on a 3.2 GHz Pentium IV processor with 1G physical memory.

### 4.3 Large CRM datasets

The thin-tail algorithm has been tested on the current version 3 of the *REDfly* database [18], which contains 894 experimentally-verified CRMs from *Drosophila*. Results show that 72.5% CRMs has $E < 0.6$ and 70.8% CRMs has $T_{50} < 0$ passing the thin-tail test. It is worth to mention to the point that the fluffy-tail algorithm has never been tested on the large CRM datasets.

## 5. Conclusions

In the thin-tail test, the statistical properties of CRMs are investigated by examining *Z-score* distribution pattern. The special statistical method used for calculating *Z-score* can reduce the effect of poly N and other simple strings on the distribution pattern of similar-word sets. Results show that the *Z-score* distribution of CRMs tends to be a thin-tail distribution as compared with that of NCNRs. Based on this observation, two thinness coefficients $E$ and $T_r$ are introduced here. By using $E$ and $T_r$, the thin-tail test has the better separation accuracy of distinguishing CRMs from NCNRs than the fluffy-tail test [9]. Especially by using the first thinness coefficient $E$, the computational time is significantly decreased, in view of a bootstrapping procedure to be required for calculating $T_r$ and $F_r$. For the example as $r = 50$, the thin-tail test with $E$ is 50 times faster than the thin-tail test with $T_{50}$, and is 6 times faster than the fluffy-tail test with $F_{50}$. Thus, the novel thin-tail test greatly simplifies the function prediction of an original input DNA sequence and can guide future experiments aimed at finding new CRMs in the post-genome time [19-23].

## References

[1] Frith MC, Li MC, Weng ZP: **Cluster-Buster: Finding dense clusters of motifs in DNA sequences**. *Nucleic Acids Research* 2003, **31**(13): 3666-3668.

[2] Bailey TL, Noble WS: **Searching for statistically significant regulatory modules**. *Bioinformatics* 2003, **19**(2): II16-II25.

[3] van Helden J, André B, Collado-Vides J: **Extracting regulatory sites from the upstream region of yeast genes by computational analysis of**

oligonucleotide frequencies. *Journal of Molecular Biology* 1998, **281**(5): 827-842.

[4]   Grad YH, Roth FP, Halfon MS, Church GM: **Prediction of similarly acting *cis*-regulatory modules by subsequence profiling and comparative genomics in *Drosophila melanogaster* and *D.pseudoobscura***. *Bioinformatics* 2004, **20**(16): 2738-2750.

[5]   Boffelli D, McAuliffe J, Ovcharenko D, Lewis KD, Ovcharenko I, Pachter L, Rubin EM: **Phylogenetic shadowing of primate sequences to find functional regions of the human genome**. *Science* 2003, **299**(5611): 1391-1394.

[6]   Emberly E, Rajewsky N, Siggia ED: **Conservation of regulatory elements between two species of *Drosophila***. *BMC Bioinformatics* 2003, **4**(57).

[7]   Li L, Zhu Q, He, X, Sinha S, Halfon MS: **Large-scale analysis of transcriptional *cis*-regulatory modules reveals both common features and distinct subclasses**. *Genome Biology* 2007, **8**(6): R101.

[8]   Nazina AG, Papatsenko DA: **Statistical extraction of *Drosophila cis*-regulatory modules using exhaustive assessment of local word frequency**. *BMC Bioinformatics* 2003, **4**(65).

[9]   Abnizova I, te Boekhorst R, Walter K, Gilks WR: **Some statistical properties of regulatory DNA sequences, and their use in predicting regulatory regions in the *Drosophila* genome: The fluffy-tail test**. *BMC Bioinformatics* 2005, **6**(109).

[10]  Chan BY, Kibler D: **Using hexamers to predict *cis*-regulatory motifs in *Drosophila***. *BMC Bioinformatics* 2005, **6**(262).

[11] Kantorovitz MR, Kazemian M, Kinston S, Miranda-Saavedra D, Zhu Q, Robinson GE, Göttgens B, Halfon MS, Sinha S: **Motif-blind, genome-wide discovery of *cis*-regulatory modules in *Drosophila* and mouse**. *Developmental Cell* 2009, **17**(4): 568-579.

[12] Shu J-J, Li Y: **A statistical fat-tail test of predicting regulatory regions in the *Drosophila* genome**. *Computers in Biology and Medicine* 2012, **42**(9): 935-941.

[13] Su J, Teichmann SA, Down TA: **Assessing computational methods of *cis*-regulatory module prediction**. *PLoS Computational Biology* 2010, **6**(12), 1001020.

[14] Papatsenko DA, Makeev VJ, Lifanov AP, Régnier M, Nazina AG, Desplan C: **Extraction of functional binding sites from unique regulatory regions: The *Drosophila* early developmental enhancers**. *Genome Research* 2002, **12**(3): 470-481.

[15] Wingender E, Chen X, Fricke E, Geffers R, Hehl R, Liebich I, Krull M, Matys V, Michael H, Ohnhäuser R, Prüβ M, Schacherer F, Thiele S, Urbach S: **The TRANSFAC system on gene expression regulation**. *Nucleic Acids Research* 2001, **29**(1): 281-283.

[16] Leung MY, Marsh GM, Speed, TP: **Over- and underrepresentation of short DNA words in herpesvirus genomes**. *Journal of Computational Biology* 1996, **3**(3): 345-360.

[17] Régnier M: **A unified approach to word occurrence probabilities**. *Discrete Applied Mathematics* 2000, **104**(1-3): 259-280.

[18] Gallo SM, Gerrard DT, Miner D, Simich M, Des Soye B, Bergman CM, Halfon MS: *REDfly* **v3.0: Toward a comprehensive database of**

**transcriptional regulatory elements in *Drosophila***. *Nucleic Acids Research* 2011, **39**(1): D118-D123.

[19]  Shu J-J, Ouw LS: **Pairwise alignment of the DNA sequence using hypercomplex number representation**. *Bulletin of Mathematical Biology* 2004, **66**(5): 1423-1438.

[20]  Shu J-J, Li Y: **Hypercomplex cross-correlation of DNA sequences**. *Journal of Biological Systems* 2010, **18**(4): 711-725.

[21]  Shu J-J, Wang Q-W, Yong K-Y: **DNA-based computing of strategic assignment problems**. *Physical Review Letters* 2011, **106**(18): 188702.

[22]  Shu J-J, Yong KY, Chan WK: **An improved scoring matrix for multiple sequence alignment**. *Mathematical Problems in Engineering* 2012, **2012**(490649): 1-9.

[23]  Shu J-J, Yong KY: **Identifying DNA motifs based on match and mismatch alignment information**. *Journal of Mathematical Chemistry* 2013, **51**: 1-9.

**Table Captions**

Table 1:   Classification of 120 sequences

Table 2:   Evaluation of $E$, $T_{50}$ and $F_{50}$

# Tables

### Table 1. Classification of 120 sequences

#### (a) Thin-tail test with $E$

| Functional type | $E < 0.6$ | $E > 0.6$ | Positive rate | Negative rate |
|---|---|---|---|---|
| CRMs | 43 | 17 | 71.7% | 28.3% |
| NCNRs | 25 | 35 | 41.7% | 58.3% |

#### (b) Thin-tail test with $T_{50}$

| Functional type | $T_{50} < 0$ | $T_{50} > 0$ | Positive rate | Negative rate |
|---|---|---|---|---|
| CRMs | 44 | 16 | 73.3% | 26.7% |
| NCNRs | 24 | 36 | 40% | 60% |

#### (c) Fluffy-tail test

| Functional type | $F_{50} > 2$ | $F_{50} < 2$ | Positive rate | Negative rate |
|---|---|---|---|---|
| CRMs | 49 | 11 | 81.7% | 18.3% |
| NCNRs | 31 | 29 | 51.7% | 48.3% |

**Table 2.** Evaluation of $E$, $T_{50}$ and $F_{50}$

**(a) Distinguish CRMs from NCNRs**

|  | The thin-tail test | | The fluffy-tail test |
|---|---|---|---|
|  | $E$ | $T_{50}$ | $F_{50}$ |
| SN | 71.7% | 73.3% | 81.7% |
| SP | 58.3% | 60% | 48.3% |
| Accuracy | 65% | 66.7% | 65% |

**(b) CPU time for a sequence length of 1000**

|  | The thin-tail test | | The fluffy-tail test |
|---|---|---|---|
|  | $E$ | $T_{50}$ | $F_{50}$ |
| CPU time | 54 second | 2700 second | 310 second |

**Figure Captions**

Figure 1:   A flow chart of thin-tail test

Figure 2:   Histogram of CRMs ($m=5$, $j=1$, $k=-0.3$)

Figure 3:   Histogram of CRMs ($m=5$, $j=1$, $k=-0.14$) after random shuffle

Figure 4:   Histogram of NCNRs ($m=5$, $j=1$, $k=0.54$)

Figure 5:   Histogram of NCNRs ($m=5$, $j=1$, $k=0.15$) after random shuffle

Figure 6:   Histograms for CRMs and NCNRs classified by $E$ ($m=5$, $j=1$)

Figure 7:   Histograms for CRMs and NCNRs classified by $T_{50}$ ($m=5$, $j=1$)

# Figures

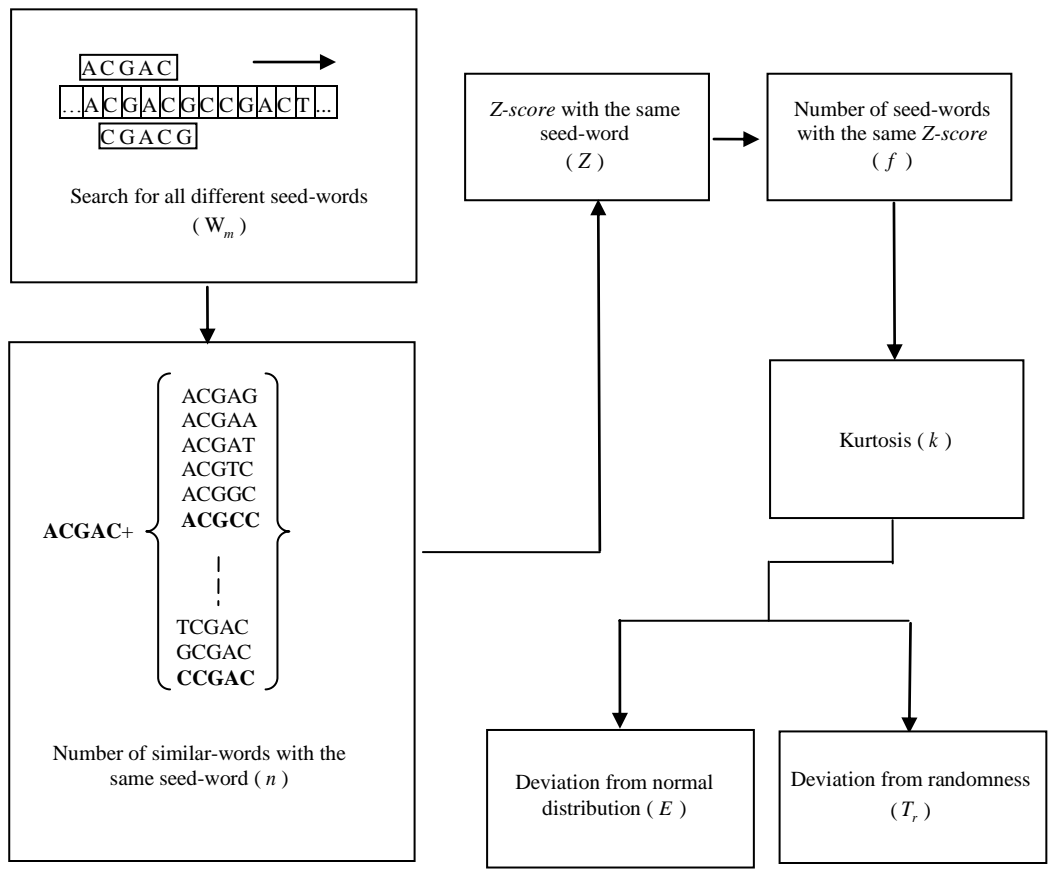

**Figure 1**     A flow chart of thin-tail test

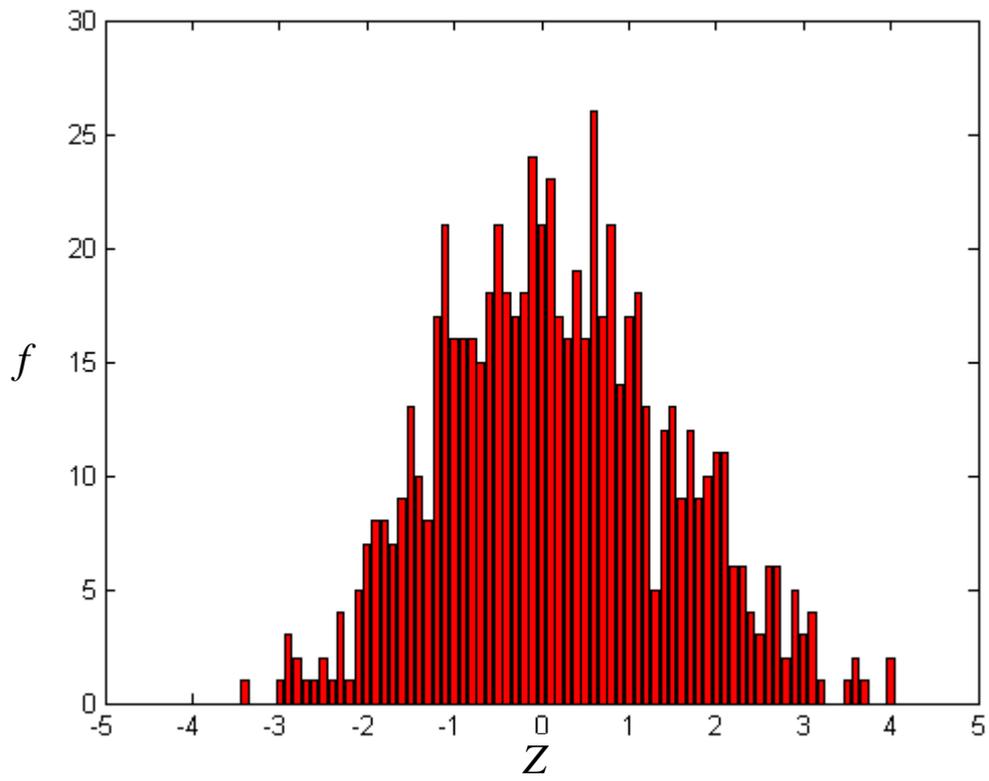

**Figure 2**     Histogram of CRMs ($m = 5$, $j = 1$, $k = -0.3$)

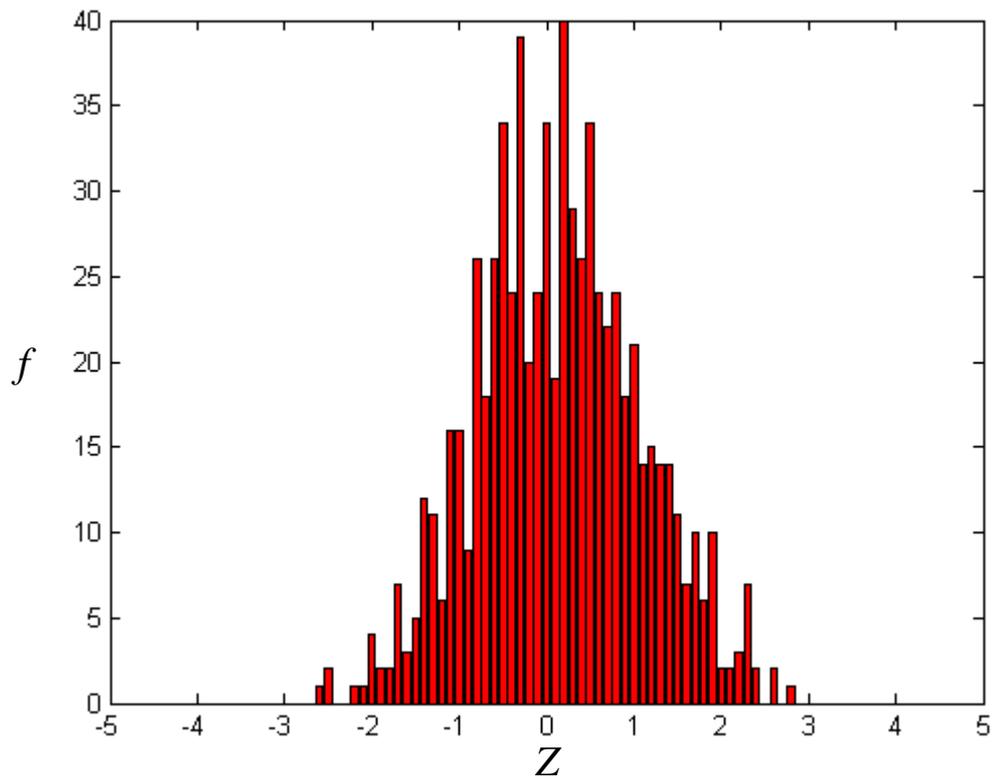

**Figure 3**   Histogram of CRMs ($m = 5$, $j = 1$, $k = -0.14$) after random shuffle

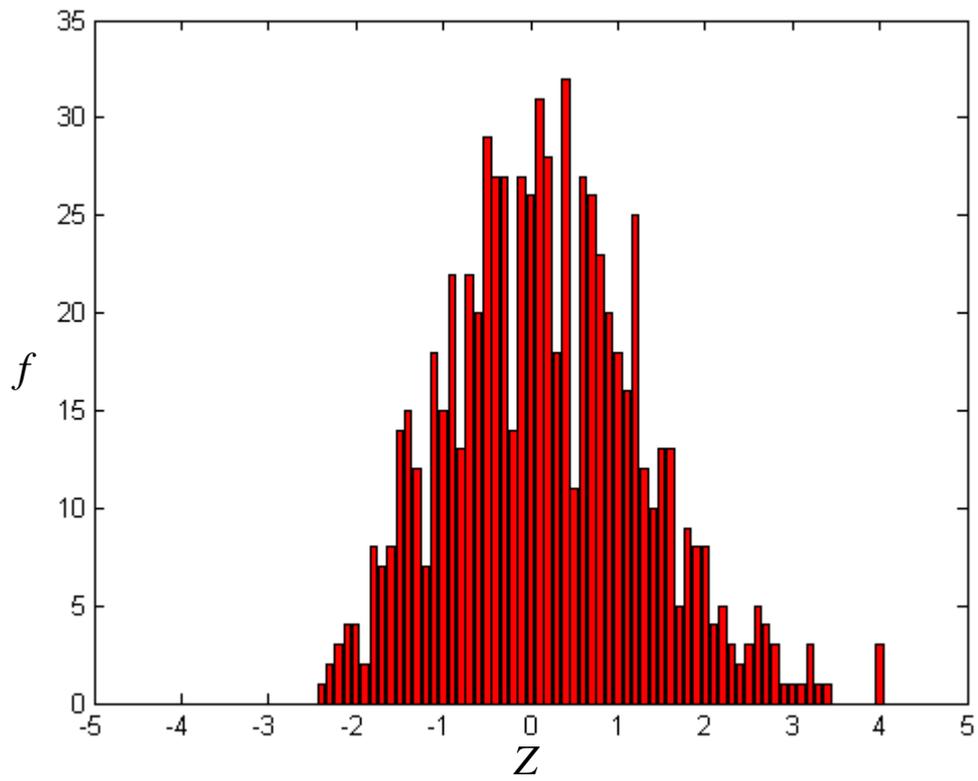

**Figure 4**     Histogram of NCNRs ( *m = 5*, *j = 1*, *k = 0.54* )

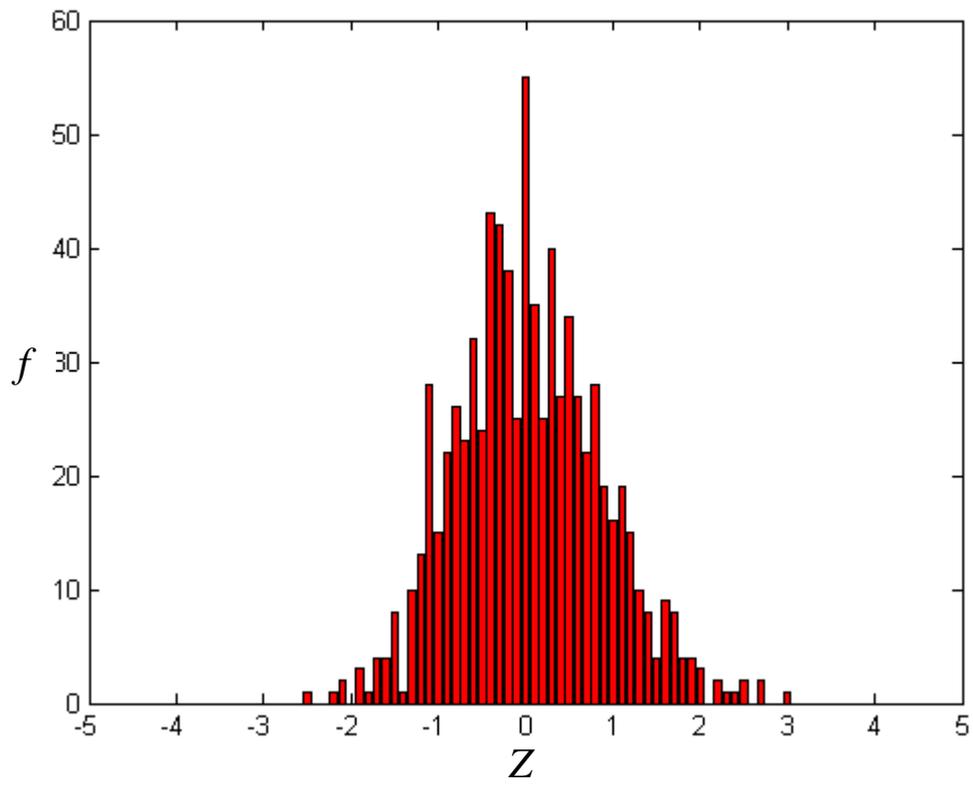

**Figure 5**　　Histogram of NCNRs ($m = 5$, $j = 1$, $k = 0.15$) after random shuffle

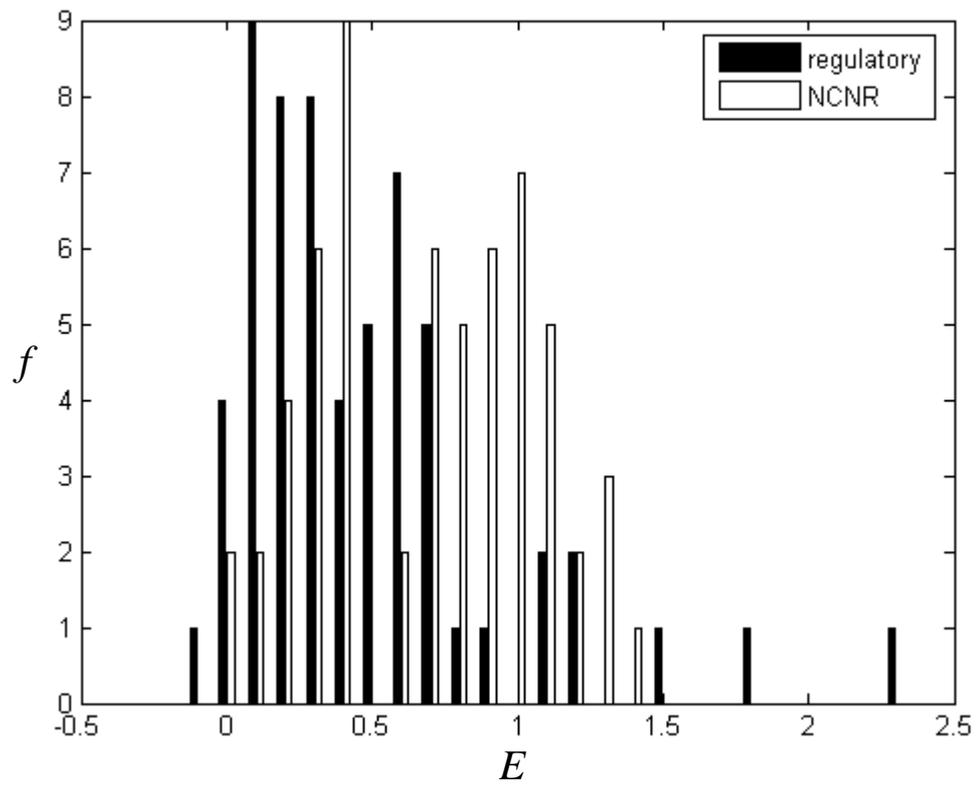

**Figure 6**   Histograms for CRMs and NCNRs classified by $E$ ($m = 5$, $j = 1$)

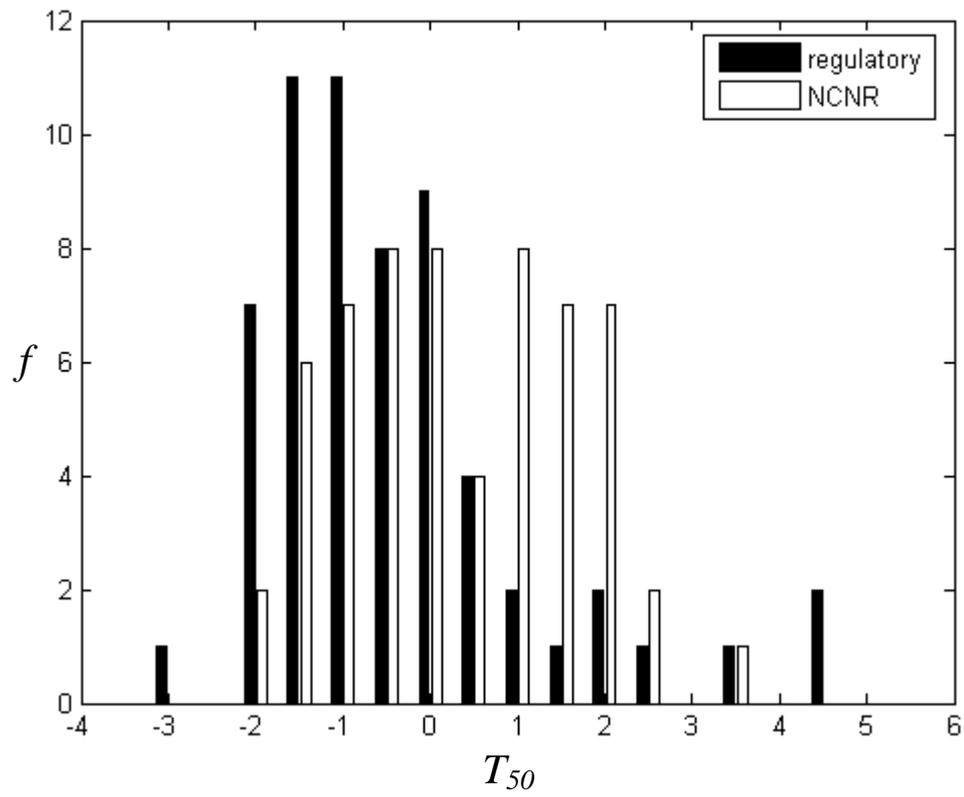

**Figure 7**     Histograms for CRMs and NCNRs classified by $T_{50}$ ($m = 5$, $j = 1$)